\documentclass [prl,amsmath,twocolumn]{revtex4-1}
\usepackage{mathptmx}
\usepackage{microtype}

\usepackage{color}
\usepackage{ifpdf}
 \ifpdf
  \usepackage[pdftex]{graphicx}
 \else
  \usepackage{graphicx}
\fi
\usepackage{color}

\usepackage{dcolumn}
\usepackage{amsmath}
\usepackage{mathtools}
\usepackage{bm}
\usepackage{amssymb}

\usepackage{mathptmx}
\usepackage{siunitx}

\RequirePackage{xspace}

\begin{document}

\title{General Relativistic Wormhole Connections from Planck-Scales and the ER = EPR Conjecture}


\author{Fabrizio Tamburini}
\email[E-mail: ]{fabrizio.tamburini@gmail.com}
\affiliation{ $^{1}$* ZKM---Zentrum f\"ur Kunst und Medientechnologie, Lorentzstr. 19, D-76135 Karlsruhe, Germany}

\author{Ignazio Licata}
\email[E-mail: ]{ignazio.licata3@gmail.com} 
\affiliation{Institute for Scientific Methodology (ISEM), Via Ugo La Malfa 153, Palermo, I-90146, Italy}
\affiliation{School of Advanced International Studies on Theoretical and Nonlinear Methodologies of Physics, I-70124~Bari, Italy}
\affiliation{International Institute for Applicable Mathematics and Information Sciences (IIAMIS), B.M. Birla Science Centre, Adarsh Nagar, Hyderabad 500 463, India}




\begin{abstract}
Einstein's equations of general relativity (GR) can describe the connection between events within a given hypervolume of size $L$ larger than the Planck length $L_P$ in terms of wormhole connections where metric fluctuations give rise to an indetermination relationship that involves the Riemann curvature tensor. At low energies (when $L \gg L_P$), these connections behave like an exchange of a virtual graviton with wavelength $\lambda_G=L$ as if gravitation were an emergent physical property. Down to Planck scales, wormholes avoid the gravitational collapse and any superposition of events or space--times become indistinguishable. 
These properties of Einstein's equations can find connections with the novel picture of quantum gravity (QG) known as the ``Einstein--Rosen (ER)=Einstein--Podolski--Rosen (EPR)'' (ER = EPR) conjecture proposed by Susskind and Maldacena in Anti-de-Sitter (AdS) space--times in their equivalence with conformal field theories (CFTs). 
In this scenario, non-traversable wormhole connections of two or more distant events in space--time through Einstein--Rosen (ER) wormholes that are solutions of the equations of GR, are supposed to be equivalent to events connected with non-local Einstein--Podolski--Rosen (EPR) entangled states that instead belong to the language of quantum mechanics. Our findings suggest that if the ER = EPR conjecture is valid, it can be extended to other different types of space--times and that gravity and space--time could be emergent physical quantities if the exchange of a virtual graviton between events can be considered connected by ER wormholes equivalent to entanglement connections.
\end{abstract}


\maketitle

\section{Introduction}
The formulation of an effective theory of quantum gravity can be considered the holy grail of modern physics. Gravitation was the first force to be mathematically described by Newton and it is the last force of nature that has yet to be quantized. 
As pointed out by DeWitt in his early pioneering works \cite{dewitt1,dewitt2,dewitt3,dewitt4,dewitt5}, since the introduction of quantum field theory  around 1930 by Heisenberg, Dirac, Pauli, Fock, Jordan and others,  many attempts were made  to find a robust and logically closed method of quantizing the gravitational field, without success, even if  Einstein's equations are known to remain valid down to the Planck scales.
Rosenfeld \cite{rosenfeld, rosenfeld2} realized the difficulty of finding general methods to quantize gravity and that the quanta of the field, if they do exist, cannot give observational effects until reaching a very high energy $E_p=\sqrt{\hbar c^5/G}\simeq 1.22 \times 10^{19}$ GeV that corresponds to the so-called Planck length, $L_p=\sqrt{\hbar G/c^2}$.
Thus, Planck scales were somehow ``artificially'' introduced in the framework of general relativity (GR), a classical theory, in the attempt of building a quantum theory of gravitation based on the finite quantum of action $h$ linked with the gravitational constant $G$ and the speed of light~$c$. 

In principle, following the works by Pauli, De Witt and other pioneers in this field, the fundamental building block for a quantum theory of gravity is the graviton, a spin-$2$ massless particle with the well-known limitations in the building of a QG theory due to the coupling constant of the gravitational field that depends on the inverse square of the mass. This coupling constant makes the Einstein--Hilbert Lagrangian of quantum gravity  divergent at the loop level. It is a non-renormalizable theory, unless introducing additional concepts such as supersymmetry like in string theory and supergravity. Up~to now no supersymmetric partners
of the known quanta have been found from the Large Hadron Collider and other experiments presented by the Particle Data Group \cite{pdata}. At the present moment one can consider different approaches, including string theory scenarios that do not require supersymmetric partners at the explored energies, or to consider a model of the Universe without strings, or adopt the approach of loop quantum gravity \cite{rovelli}, where gravitons do not represent the building blocks of the theory. 
The interactions between events that can be ascribed to graviton exchanges can be recovered in a weak field limit approximation.
The exchange of a virtual graviton between two particles does not have the support of an actual theory of quantum gravity. As an example, in string theory and in most quantum field theory (QFT) 
 scenarios, in the building of the theory, one must introduce the quanta of the associated field. The~quanta are introduced in terms of quantized excitations on a classically fixed background.
The main conceptual problem for the formulation of a consistent theory of QG is that this theory must unify and contain as special cases both GR and quantum mechanics (QM). Unfortunately, GR has concepts and mathematical structures that are incompatible with those of  QM and vice versa, with the result that the two theories do not communicate between each other.
GR is a local deterministic theory based on point-to-point connections of events and observers that define a four-dimensional manifold. Einstein in 1947, his latest memoirs,  
stated that space--time is made with connections between events and, more precisely, with coincidences of events. On the other hand, QM presents non-locality and the well-known probabilistic behavior from the deterministic equations that rule the quanta.

This contrast between GR and QG can find a fusion in the simple heuristic approach formulated by Susskind and Maldacena who, starting from the quantum mechanical language, set an hypothetical equivalence between non-traversable wormhole connections of two (or more) particles or events in space--time through Einstein--Rosen (ER) bridges and entangled states (the idea that wormholes and flux tubes can play a role in quantum mechanics and quantum field theory is not new, in particular for systems with electric and/or magnetic charges and their renormalization has earlier work in \cite{Dzhunushaliev,Dzhunushaliev1}),
 and the quantum properties of the ``spooky action at distance'' of Einstein--Podolski--Rosen (EPR) states  \cite{er,epr,mald13,suss16}.  The ER=EPR equivalence was first defined in Anti-de-Sitter (AdS) space--times in their equivalence with CFTs \cite{adscft1,witten,adscft2,adscft3,adscft4}.
In other words, EPR entangled particles are supposed to be equivalent to connections obtained through ER wormholes involving the concept of entanglement entropy to describe these many-body quantum state/wormhole connections, even if the ER = EPR equivalence is more evident with monogamous entangled pairs \cite{Gharibyan}.
Spacetime is supposed to emerge from quantum entanglement, as discussed in \cite{vanraamsdonk} where, from some examples where gauge theory/gravity duality is valid, one finds that the emergence of space--time is related to the quantum entanglement of the degrees of freedom present in these quantum systems. Superpositions of quantum states corresponding to disconnected space--times can give rise to states that are interpreted in terms of classically connected space--times. In this vision, gravity can be also interpreted as an entropic force, a thermodynamic property of physical systems defined in an holographic scenario: gravity and space--time connections are emergent phenomena from the degrees of freedom of a physical system encoded in an holographic boundary or to emerge 
 from a background-free approach by using quantum entanglement \cite{verlinde,cao18}.
At all effects, one can conclude that space--time is built with the quantum information shared between EPR states that are equivalently connected with an ER wormhole.

While ER wormholes are classical solutions of GR, a local deterministic theory, quantum entanglement, is instead one of the most intriguing quantum physical aspects of nature characterized by non-locality and the stochastic properties of quantum mechanics. Entanglement occurs when a pair of particles (or a group of quanta) is generated in a way that the quantum state of each particle of the pair cannot be described independently of the state of the others even if they are separated by large distances. For a deeper insight see \cite{bell,bell2,horodecki,zeilinger1,zeilinger2,zeilinger3}. 
In ER = EPR, causality is not violated. ER bridges do not violate causality because of the topological censorship, which forbids ordinary traversable wormholes; EPR states, instead, prevent causality violations because of the properties of entangled states described by Bell's inequalities---no information is transferred between the two entangled states during the wavefunction collapse of the entangled pair as each quantum state in an EPR pair cannot be described independently of the other states \cite{bell,horodecki}.
The ER = EPR equivalence is valid if there are no traversable wormhole solutions that do not require the violation of the strong and/or weak energy conditions \cite{visser1,maldacena18a,maldacena18,horowitz}, they may instead behave as quantum communication channels between the quantum fields there defined \cite{bao}.

The ER = EPR conjecture was initially formulated in the gravity/gauge theory equivalence between Anti-de Sitter (AdS) space--times and conformal field theories (CFTs) by Maldacena (gauge/gravity duality) within a relationship between the entanglement entropy of a set of black holes and the cross-section area of ER bridges connecting them.
AdS space--times represent an elegant solution of Einstein's equations with negative curvature where the outer boundary is a surface and, in the CFT, correspondence quanta can interact and generate the holographic universe there contained.
The AdS/CFT correspondence provides a complete non-perturbative definition of gravity with quantum field theory, extending this correspondence also to space--time scenarios of quantum gravity where the asymptotic behavior of the space--time is that of AdS space--time. The AdS/CFT correspondence plays a key role in the calculations of strong coupled quantum field theories. When~the boundary theory is strongly coupled, the bulk theory is weakly coupled, and vice versa. A strongly coupled field theory can have an AdS dual gravity description weakly coupled and therefore calculable and vice versa. As an example, if the curvature of the AdS space increases, the gravitational coupling becomes stronger and the boundary coupling is weaker. 

To extend ER = EPR conjecture to space--times different from the Anti-de Sitter solution, one has to investigate how much ER = EPR depends strictly on AdS/CFT correspondence and from the properties of wormholes also in de Sitter (dS) space--times.
First of all we must consider that AdS/CFT is a structural correspondence between bulk and screen, but does not contain in itself any specific indication of the possible dynamics of wormhole formation. Wormholes require a cosmological scenario. For example, it is plausible that the wormholes were formed in the initial chaotic phases of the universe with a rate similar to that of the formation of mini black holes (BHs), 
with very specific traces as regards the event horizon, as discussed in \cite{bueno,cardoso,konoply}. This aspect is decisive because all the problems related to the quantum aspect of the wormholes imply a cosmological background capable of providing a plausible scenario for their existence described by the Ryu--Takanayagi's entropy that relates the entanglement entropy in CFT and the geometry of AdS space--times. The Ryu--Takanayagi formula is a generalization of the BH entropy formula by Bekenstein--Hawking \cite{bek,haw} to a whole class of holographic theories \cite{hubney2007,fukami} where gravitational models with dimension $D$ are dual to a gauge theory in dimension $D-1$. 

In these recent years, the interest in the maximum symmetry properties of de Sitter's space gave this structure a new centrality with respect to AdS. One of the most relevant problems was to project a hologram of a quantum particle that lives in the infinite future of AdS, which makes it difficult to describe real-time space in holographic terms. In particular, the main classes of essential results must be mentioned here: the CPT Universe \cite{boyle} and the numerous results on the non-locality in dS space--times \cite{naraina,chena,mald2,narayan,arias}.
Of relevant importance is the so-called ``uplifting'' technique by Dong~et~al.~\cite{dong} where two Anti-de Sitter space--times are transformed into a de Sitter space--time. The uplifting changes the curvature of two ``saddle-shaped'' AdS space--times that, once warped, are glued together along their rims and turned into a ``bowl-shaped'' dS 
 space--time via entanglement or more general two-throated Randall--Sundrum systems \cite{Dimopoulos,Dimopoulos1} and the CFTs relative to both hemispheres become coupled with each other. In this way one forms a single quantum system that is holographically dual to the entire spherical de Sitter space, defined on its boundary located at a finite distance away.
The~technique of uplifting two AdS into a dS permits us to modify the curvature in a more general way than that offered by the set of local transformations obtained through Wick rotations that can only act locally---the curvature changes everywhere by introducing extra fields whose energy density acts as an extra source of curvature to landscape the AdS space--time into a dS one. The cosmological constant in the bulk space is then transformed from negative to positive and the holographic projection of the space--time into its boundaries is changed.
Some examples are reported by Silverstein and Polchinski~\cite{Polchinski} or in Vasiliev's higher-spin gravity---in AdS, the boundary theory is an O(N)-vector field theory, while in dS space it becomes an Sp(N) scalar field theory, where $N$ is the number of the vector (scalar) fields of the boundary theories \cite{Vasiliev,danninos}.

In this work, we analyze Einstein's equations in a finite volume of space--time down to the Planck scale, finding wormhole connections that avoid the singularity problem and an indetermination relationship that involves the Riemann curvature. This finds application to the ER = EPR conjecture---in this framework, geometry behaves as a geodesic tensor network that defines the quantum state properties of a fundamental quantum state of a given metric \cite{dongzhou} and a virtual graviton exchange becomes equivalent to entanglement to which one can apply the concept of Penrose's decoherence of a quantum state \cite{penrose}.
In this crossing between locality of GR and the emergence of non-locality of QM as in \cite{chiatti,cordafeleppa}, where de Sitter space--time is taken as the geometric structure of vacuum, the analysis of Einstein's equations can provide an additional support to
the ER = EPR conjecture extended from AdS to dS \cite{strominger} and to locally Euclidean space--times. 
This can be interpreted as the route to ER = EPR from general relativity. 

\section{Wormhole Connections down to Planck Scales from Einstein's Equations}

In the ER = EPR scenario, wormhole connections are fundamental in the building of space--time.
Consider an entangled quantum system. The emergence of space--time in terms of ER connections, in the gravity picture, is intimately related to the quantum entanglement of degrees of freedom in the corresponding conventional quantum system, building up space--time with quantum entanglement. The ER = EPR equivalence suggests that space--time and gravity may emerge from the degrees of freedom of the field theory. On the other hand, space--time becomes the optimal way to build entanglement starting from wormhole connections.
At Planck scales, the Planck area is defined as the area by which the surface of a Schwarzschild black hole increases when in the black hole is injected one bit of information.
In a Riemannian manifold $(M,g)$ the scalar curvature in an $(n-1)$ hyperplane relates GR with the entanglement of quantum states in an arbitrary Hilbert space without reference to AdS=CFT or any other holographic boundary construction.

\subsection{Einstein's Equations in the Neighborhood of an Event}
Einstein's equations are the core of GR---they describe gravity in terms of the curvature of space--time. Spacetime geometry and the metric tensor $g_{ik}$ are determined from Einstein's equations, given the distribution of energy, mass and momentum in space--time encoded in the stress--energy tensor $T_{ik}$.

In our approach, by assuming that Einstein's equations remain valid down to the Planck scale, we find that the connection between events are achieved through wormhole connections, avoiding the gravitational collapse and the presence of singularities at Planck scales. To this aim, we adopt the approach by Schwinger in the analysis of classical fields \cite{schwinger}---to determine the properties of a field, one cannot measure the field in a point, otherwise, because of the equivalence principle, one finds only a local Minkowskian space--time tangent to the manifold $(M,g)$ in the given point event.

Following Schwinger, from his studies on electromagnetic theory \cite{schwinger,thidebook}, the analysis of a classical field must be made in a neighborhood of the event. This approach is clearly valid in electromagnetism and antenna theory: when sensing the electromagnetic field with an antenna, one cannot measure in a point the fluctuations of the electric field. The antenna must have a finite length in space and the field must be measured in a finite time interval to be revealed.
As happens for any antenna, as well as for the gravitational field, one has to consider a finite length or a finite hypervolume in which to determine the properties of the field.
For the same reasons, one cannot deduce the properties of the gravitational field in a single  point and at a given time because of the equivalence principle, as discussed in the free-falling particle paradox \cite{falling}.

Let us find the main properties of the gravitational field when a finite length $L$ of measure is fixed down to the Planck scale.
Given a Riemannian manifold $(M,g)$, where $M$ is the manifold and $g$ the metric tensor, $g \in \otimes^2 \dot T$ of  tensorial order $2$, (written as $g_{(2)}$), Einstein's equations are
\begin{equation}
R_{ik}-\frac 12 R g_{ik} + \Lambda g_{ik}=8\pi T_{ik}
\end{equation}
where $\Lambda$ is the cosmological constant and $R_{ik}=g^{lm}R_{lmik}$ and $R=g^{ik}g^{lm}R_{lmik}$ represent the tensorial and scalar space--time curvature terms obtained from the Riemann tensor $R_{iklm}$ that, in the tensorial index notation, takes the usual well-known form \cite{landau2}
\begin{eqnarray}
R_{iklm}&=&\frac12 \left(\frac{\partial^2 g_{lm}}{\partial x^k \partial x^l } + \frac{\partial^2 g_{kl}}{\partial x^i \partial x^m } - \frac{\partial^2 g_{il}}{\partial x^k \partial x^m } - \frac{\partial^2 g_{km}}{\partial x^i \partial x^l }\right)+  \nonumber
\\
&+& g_{np}\left(\Gamma^n_{kl}\Gamma^p_{im} - \Gamma^n_{km}\Gamma^p_{il}\right)
\end{eqnarray}
the tensor is an element of the rank-four tensors $R_{iklm} \in \otimes^4 \dot T$ in the cotangent bundle $\dot T$ of the manifold $(M,g)$ that we will indicate with the symbol $R_{(4)}$, where $4$ is the tensorial index. 

The gravitational field has the fundamental property that, any body, independently from their mass, moves in the same way. This is described by the strong equivalence principle, which suggests that gravity is a geometrical quantity and one cannot measure the gravitational field in an event, as the field becomes locally Galilean and diagonalizable, and the energy of the field cannot be uniquely~defined. 

In a neighborhood of a given event, space--time is built by chains of events and observers. The~building of space--time is obtained by causally transferring information encoded locally in one event of the field to form events and coincidences of events described by punctual (point-to-point) correlations 
in a four-dimensional manifold $(M,g)$.

Because of the equivalence principle, these observables must be generated within a volume of finite spatial extent from a given spatial length $L$ and propagated---through a chain of events---in the form of four-volumetric densities (or in geometrical sub-varieties) in the four-dimensional manifold $(M,g)$ to a remotely located finite region of space--time of likewise finite spatial/temporal extent---the observation (hyper-)volume $V\subseteq M$ over which they are volume integrated into observables, allowing the information carried by them to be extracted and decoded. More specifically, we will use the local split of $3+1$ in space and time where we will consider the integration of the field properties over a three-dimensional volume $V=L^3$.
The volumetric density of every gravitational observable carried by the gravitational field is a linear combination of quantities that are second order (quadratic/bilinear) in the metric of the field, and/or of the derivatives of the field.
To obtain these quantities one must integrate the density of a conserved quantity e.g., over a given 3D space-like hypersurface $\sigma$ or over a four-dimensional interval, characterized by a given finite length $L$; there, the field equations are integrated and averaged to obtain the field observables we need to analyze the properties of space--time down to the Planck scales.

Let us consider a Lorentzian manifold as example, with cosmological constant $\Lambda$ and then introduce a characteristic length $L$. This quantity is the proper length associated to the generic coordinate variation $\Delta x$ written in terms of the metric tensor $g$, viz.,  $L \sim g^{1/2} \Delta x$. The Riemann tensor is then written in terms of the metric variations $\Delta g$, the covariant metric tensor $g$ and the contrarvariant one, $g^{-1}$

\begin{equation}
R_{(4)}(g,L) \sim \frac{g^2}{L^2} \left( \Delta\left(\Delta g (g)^{-1}\right) + \left(\Delta g (g)^{-1}\right)^2\right)
\end{equation}
where the term $\left(\Delta g~ (g)^{-1}\right)^2_{\{kl,im-km,il\}} = \Gamma^n_{kl}\Gamma^p_{im} - \Gamma^n_{km}\Gamma^p_{il}$  represents the affine connection and $\Delta\left(g^{-1}\Delta g\right)$ is the second derivatives of the metric tensor with respect to the coordinates, being $\partial^2_{kl}~g = \frac{g}{L^2}\Delta_k\Delta_l~g = 
\frac{g^2}{L^2} \Delta_k \left(\Delta_l g (g)^{-1}\right)$.
The Ricci tensor and scalar are $R_{(2)}\sim ~ g^{-1}R_{(4)}$ and $R\sim ~ g^{-2}R_{(4)}$, the Einstein tensor is $G = R_{(4)}(g)^{-1}$ and Einstein's equations are $G+g\Lambda=T$, where $T$ is the energy--momentum tensor.

By introducing a characteristic length $L$, if Einstein's equations hold down to the Planck scales, from the basic formulation of the Riemann tensor and Einstein equations we find that the field equations, integrated and averaged over a 3D space-like hypersurface $\sigma$ with unit normal vector $n\sim g^{-1/2}$, obey an indetermination relationship that recalls Heisenberg's.
Instead of focusing on the more general energy--tensor quantity (or the momentum vector), we consider for the sake of simplicity the scalar proper energy $E$, averaged over a proper volume $L^3$, which is given by the integral of the energy momentum tensor over a given proper volume element of a space-like 3D--hypersurface.
This~leads to the following formulation of the proper energy averaged over the given proper volume
\begin{equation}
\left< E - g\Lambda \right> = \bar E\sim \frac{g^2}{L}R_{(4)}= L \left( \Delta\left(\Delta g (g)^{-1}\right) + \left(\Delta g (g)^{-1}\right)^2\right).
\end{equation}

If we rescale this relationship down to the Planck scale $L_p$, by defining the light crossing time as $\tau = L$ and the Planck Time $\tau_p$, the Einstein equations retain their validity down to the Planck scale, even if metric fluctuations over a scale larger than $L_p$ can occur.
We find that these fluctuations can give rise to a relationship
\begin{equation}
\left(\frac{\tau_p}{\tau}\right)^2 \left(\frac{\bar E\times \tau}{\hbar}\right)=\left(\frac{L_p}{L}\right)^2 \left(\frac{\bar E\times \tau}{\hbar}\right)=\frac{L^2}{g^2}R_{(4)}(g,L)
\label{fluttua}
\end{equation}
that holds down to the Planck scales.
Fixing a characteristic spatial scale (or time), the relationship in Equation \eqref{fluttua} corresponds to the introduction of fluctuations of the averaged quantity over $L^3$ of the proper energy $\bar E$. If we set $\bar E = \Delta E^*$ and $\tau = \Delta t$, we can write Equation \eqref{fluttua} in a more familiar Heisenberg relationship that involves the Riemann tensor and the contribution from the dark energy
\begin{equation}
 \Delta E^*\times \Delta t=\hbar \left(\frac{\tau}{\tau_p}\right)^2\frac{L^2}{g^2}R_{(4)}(g,L)=\frac{\hbar}{g^2} \left(\frac{L^2}{L_p}\right)^2 R_{(4)}(g,L)
 \label{ind1}
\end{equation}
that at Planck scales becomes
\begin{equation}
 \Delta E^*\times \Delta t=\hbar \frac{L_p^2}{g^2}R_{(4)}(g,L)=\hbar \left( \Delta\left(\Delta g (g)^{-1}\right) + \left(\Delta g (g)^{-1}\right)^2\right)
\label{ind2}
\end{equation}
where $\Delta E^* = \Delta E + \Delta g ~ \Lambda + g \Delta \Lambda$ averaged on the volume $L^3$ of the 3D space-like hypersurface $\sigma$.

\subsection{The Energy of the Gravitational Field}
Dark energy and other different vacua are parameterized by the cosmological constant $\Lambda$.
When $\Lambda>1$, the equations describe an AdS space--time. The gravitational fluctuations are mainly expressed by the affine connection term $\left(\Delta g (g)^{-1}\right)^2$ for any space--time. 
To describe the energy of the gravitational field, which is not defined as a global and conserved quantity, one has to introduce pseudotensorial quantities describing the energy trapped non-locally in the geometry.
One example is the non-symmetric Einstein pseudotensor, which is constructed exclusively from the metric tensor and its first derivatives but is not suitable for our purposes. Instead, the Landau--Lifshitz pseudotensor $t_{ik}$ \cite{landau2} permits us to write for the integrated non-local gravitational energy $E_g$ that includes the contribution of the cosmological constant in terms of the curvature tensor. This quantity is quadratic in the connection and, for a general covariant component of the pseudotensor,  averaged on the volume $V=L^3$ one obtains
\begin{equation}
\left< g^{-1} \left(t +\Lambda g\right)\right>_{V=L^3} \sim \frac{E_g+E_\Lambda}{L^3}
\end{equation}
where $E_\Lambda= g^{-1} \Lambda g L^3$ is the energy associated to the value of the cosmological constant and to dark energy. Considering that
\begin{equation}
|g| ~g^{-2} \left(t +\Lambda g\right)\sim \left(\frac{\Delta g}{\Delta x}\right)^2,
\end{equation}
this relationship leads to a background curvature with fluctuations having wavelength $\lambda=L$ that can be interpreted as connections between events due to an exchange of virtual gravitons with wavelength $\lambda$ and energy $\hbar/\lambda$ or, in the ER = EPR scenario, to the connection through an ER wormhole,
\begin{equation}
\left<\left(\Delta g (g)^{-1}\right)^2\right>_{V=L^3} \sim \frac{E_g+E_\Lambda}{L}=\left(\frac{\tau_p}{\tau}\right)\frac{E_g+E_\Lambda}{ E_p}
\label{fluttualarge}
\end{equation}
where $E_p$ is Planck's energy. In the ER = EPR hypothesis, these energy fluctuations would be considered as equivalent to the connection between two entangled events separated by the distance $L$, giving a paradoxical meaning to the exchange of a virtual graviton in terms of entanglement connections between events like in an emergent gravity scenario.

Following the already cited works by De Witt and the classical QG interpretation found in the literature \cite{dewitt1,dewitt2,dewitt3,dewitt4,dewitt5}, this term would describe a virtual graviton exchange between two events within a space--time connection. On the other hand, this term---that can be also interpreted in terms of a wormhole connection between the two events---with the exchange of at least 1 qbit of information (in the ER = EPR conjecture) would correspond to the entanglement of two particles.
If ER wormholes are equivalent to a monogamous connection between the two events \cite{Gharibyan}, as realized through a virtual graviton exchange, one could state that entanglement of EPR states can derive from the exchanges of virtual gravitons between two events. From another perspective, entangled states should be provided by a mixed state between the two entangled pairs with that of the virtual graviton. The question is what is the correct perspective?

From Einstein's equations, the observable averaged total energy of a metric fluctuation over the volume $V$ on a scale $L$, becomes
\begin{equation} 
E\sim \left(\frac{\tau}{\tau_p}\right)\left< \Delta \left(\Delta g (g)^{-1}\right)\right>_{V}E_p + E_g+E_\Lambda
\end{equation}
and is made with the energy of geometry and vacuum and energy of interaction expressed in terms of gradients of the geometry fluctuations, second order derivatives of the metric tensor, as in the Riemann tensor that make the connection between observers.

\subsection{Planck-Scale Wormhole Connections}
In a local neighborhood of a given event $\{x^i\}_0$, one performs a discrete infinite denumerable $3+1$ local slicing of the space--time with time steps a Planck time unit.
To the initial event $\{x^i\}_0$ corresponds the slice $N=0$. The $N-$th slice corresponds to the time $\tau_N = (N+1)\tau_p$, building up a symbolic dynamics of space--time events.
The energy of the gravitational perturbation in the $N-$th slice is 
\begin{equation}
\left(E_g+E_\Lambda \right)_N \sim \frac{E_p}{N+1}
\end{equation}
the total energy is instead
\begin{equation}
E_N \sim \left(N+1\right) \left<\Delta  \left(\Delta g (g)^{-1}\right)\right>_{V,N} E_p + \frac{E_p}{N+1}
\end{equation}
for $N\rightarrow 0$ the energy fluctuation becomes $E_N \rightarrow E_p$ for which, by definition, $E_N \tau_N \sim \hbar$ and the metric tidal fluctuations tend to zero---space--time at Planck lengths is homogeneous and isotropic and the local geometry depends only on the energy of fluctuations in space--time and from the energy of the cosmological constant.
This because $\Delta g/g \rightarrow 1$ are both on the order of the Planck scale. This is the reason why the field does not diverge and no singularities are present.

For $N\rightarrow \infty$ the dominant energy is that of tidal fluctuations at scales larger than $L_p$ accompanied with that of the cosmological constant when integrated over the metric and remains as a constant function over the volume of integration; $E_g$ becomes instead negligible.
This means that for a process connecting two events lasting a time $\tau$, the amount of energy does not entirely contribute to the vacuum energy but it is partially spent in geometry in this process, involving the dark energy contribution expressed by the cosmological constant $\Lambda$.
Recalling Heisenberg principle from Equation \eqref{fluttua}, the larger is the energy fluctuation, the smaller results the space/time interval fluctuation.

In the neighborhood of Planck scales, when $N>0$, the curvature of space--time remains finite and the Riemann tensor can be written as
\begin{equation}
R_{(4)}(g,L) \sim \frac{E_p}{\hbar}\left(\frac{\tau_p}{\tau}\right)^2 \frac{g^2}{L^2}
\end{equation}
and the Ricci scalar is 
\begin{equation}
R(g,L) \sim \frac{1}{L^2}\frac{E_p}{\hbar}\left(\frac{\tau_p}{\tau}\right)^2 
\end{equation}
that for $L\rightarrow L_p$ we have $R(g,L) \rightarrow 1/L_p^2$ with the result that at Planck scales there is no singularity in the curvature and the gravitational radius becomes
\begin{equation}
R_g=2 \frac{E~\tau~L_p^2}{\hbar~L}
\end{equation}
that is written as $R_g=2L_p$, which corresponds to elementary wormhole connections at the Planck scale and finding a trivial  equivalence with the corresponding Penrose diagrams.
Directly from Einstein's equations we find that, at Planck scales, the singularities expected from quantum gravity can be  interpreted in terms of wormhole connections between the events, as required in the ER = EPR conjecture and obtain an indetermination relationship shown in Equation \eqref{ind1} involving the Riemann tensor and geometry fluctuations.
In this view, wormhole and equivalent EPR connections can also be formally equivalent to an exchange of a virtual graviton at scales larger than Planck scale, whilst any group of superimposed states below Planck scales, instead, will be indistinguishable and therefore~entangled.

\subsection{Tests for the ER = EPR Conjecture} 
If we suppose the validity of the ER = EPR conjecture, the geometry fluctuations present at Planck scales may be revealed with quantum entanglement. By applying the indetermination relationship that involves the Riemann tensor expressed in Equations \eqref{ind1} and \eqref{ind2}, we argue that one can obtain information about the fluctuations of space--time and determine whether a characteristic scale like the Planck's one is present, as expected in QG.

If space--time is discrete, its discreteness is expected to be characterized by a typical scale of space and/or time: There exist a minimum time interval $t_p$ and a minimum length $L_p$ where wormholes connections---equivalent to entangled states between two or more regions of space--time---connect different events or space--times. If events/space--times are connected with intervals smaller than $L_p$ and $t_p$, they would be entangled and actually be the same event or the same space--time.
Their~quantum superposition can exist and collapse after a finite time interval and the properties of wormhole connections are reflected in the properties of the corresponding EPR states also when they connect events at scales larger than the Planck scale. This scenario is different from Penrose's assumptions \cite{penrose}, where space--time is thought to be continuous and the quantum superposition of space--times result unfeasible leading to the gravitational collapse.

Noe we propose to test the ER = EPR scenario by using the Heisenberg uncertainty principle applied to pairs (or groups) of entangled particles and including the additional indetermination introduced by quantum gravity effects.
The generalized Heisenberg's uncertainty principle for the momentum $p$ and the position $x$ that includes the existence of a characteristic length $L_p$ such as the Planck scale, or any other scale typical that can be found in certain quantum gravity models, is given by \cite{kempf,scardigli}
\begin{equation}
\Delta x = \Delta x_{QM}+\Delta x_{GR} \geq \frac{\hbar}{2\Delta p}+ k~\Delta p
\label{heisenberg}
\end{equation}
the existence of a minimum interval in space--time is revealed by a deviation from the classical term due to quantum mechanics only, $\Delta x_{QM}$.
The quantum gravity term, $\Delta_{GR}$, due to the existence of a characteristic length $L_p$ and to the properties of the gravitational field, can be characterized instead by a parameter $k$, a constant characteristic of the quantum theory of gravitation here considered. 
To give an example, in a string theory scenario, $k=\alpha Y$, where $\alpha$ is the string tension and $Y$ a constant that depends on the theory. In our case, following \cite{adler,garay,gao06},  one can find that $k=2 L_p^2/\hbar$. 
From our calculations that involve the Riemann tensor, we find that $ \Delta x_{QG}= 2 E_p/\Delta E^*$ and thus $\Delta p= \hbar E_p/\Delta E^* L_p^2$, a term that includes the effects of dark energy in the term $\Delta E^*$ too.

We write now the Heisenberg relationship for sets of $N-$particle entangled states. Following~\mbox{\cite{prevedel,blado,zeng,rigolin1,rigolin2}}, consider first a couple of entangled particles with positions $x_1$ and $x_2$ and momenta $p_1$ and $p_2$, respectively. For $N=2$, the classical indetermination principle is  
\begin{equation}
\Delta (x_1, x_2)^2_{QM}=\left[\Delta (x_1)^2+\Delta (x_2)^2\right] \times \left[\Delta (p_1)^2+\Delta (p_2)^2\right]\geq \frac{\hbar^2}{4}.
\label{heisenberg2}
\end{equation}

In the simplest case, where $\Delta (x_1)=\Delta (x_2)=\Delta x_e$ and $\Delta (p_1)=\Delta (p_2)=\Delta p_e$, the uncertainty relationship becomes $(\Delta x_e)^2 (\Delta p_e)^2 \geq \hbar^2$. For $N$ identical entangled states, the extended indetermination principle becomes
\begin{equation}
(\Delta x_e)^2 (\Delta p_e)^2 \geq \frac{N^2 \hbar^2}{4}
\end{equation}
and when we include the effects of the gravitational field one obtains
\begin{equation}
\Delta x = \Delta (x_1, x_2)_{QM}+\Delta (x_1, x_2)_{GR} \geq \frac{N \hbar}{2\Delta p}+\frac{2 N  L_p^2 \Delta p}{\hbar}.
\label{heisenberg3}
\end{equation}

By assuming that Einstein's equations retain their validity down to the Planck scales and that wormhole connections represent the building blocks of the physics of the gravitational field at and below Planck scales ($t_p$ and $L_p$), ER = EPR links connecting any space--time (or event) with a difference smaller than $t_p$ and $L_p$ mean that different space--times and events are physically identical, and then in principle undetectable and entangled. Instead, in a region with radius $R$, the spatial difference of two space--times/events is $\Delta L = 2 L^2_p \Delta E^*/\hbar c$, and the difference of their corresponding space--times is the difference of the proper spatial sizes of the regions occupied by them and the time of the wavefunction collapse is on the order of $\tau_c \sim 2 \hbar E_p/(\Delta E^*)^2$.

If the properties of ER = EPR links remain valid from the Planck up to the macroscopic scales, where entanglement can be observed in the lab, the term $\Delta x_{QG}$ in the Heisenberg relationship expressed in Equation~\eqref{heisenberg3} is expected to reveal the properties of the wormhole structure of space--time from a deep analysis of the wavefunction collapse of an entangled pair.
In other words, the deviation from the quantity $\Delta x_{QM}$ of the classical Heisenberg principle would reveal the fuzziness space--time or, better, of the point-by-point identification of the spatial section of the two events/space--times, better evident with a set of a large number of $N$ entangled quanta like a Schr\"odinger cat.

From the point of view of relativistic quantum information discipline, entanglement and wormholes are expected to create space--time and entanglement events (and space--times) through quantum information---information that emerges from the connection of quantum bits. 
In fact, from a quantum-computational interpretation of space--time entanglement, in a foliation of space--time, the quantum fluctuations of the metric present on the slice $n$ can be interpreted as wormhole connections between one Planckian pixel in the slice $n$ with that one present in the $n-1$  slice. Following \cite{zizzi18}, the~holographic principle suggests that such a geometrical connection is space--time entanglement. If not entangled, following Penrose's argumentation only the quantum superposition of two space--times with a difference larger than the minimum sizes can not exist, and should collapse instantaneously. If~they are connected by an ER wormhole they should obey the indetermination relationship expressed in Equation \eqref{heisenberg3}.

To verify possible additional anomalies in the indetermination principle introduced by the ER~=~EPR conjecture one may instead want to consider to measure the time/energy entangled states and study the time of collapse as a function of their energy differences. This may explain why the wavefunction collapse of an EPR pair is not always instantaneous, as it may depend on the geometry fluctuations. Moreover, one has to also consider  the effects introduced by the presence of the cosmological constant, of the information encoded and shared between the entangled quantum states and their relationship with the gravitational information entropy that go beyond the purpose of the present work.

Of course an experimenter has to consider that EPR states depend on the choice of reference frames and that Bell's inequalities are preserved in certain reference frames only, and should also consider the effects of simultaneity and  include in the experiment the additional macroscopic effects induced by the gravitational field at large scales in the presence of massive bodies.
As an example, simultaneity is responsible for the uncertainty of the ordering of non-local wavefunction collapse when the relativistic effects cannot be neglected. In any case, if a time measurement performed with an entangled pair of photons is seen as simultaneous in one shared reference frame, then the result of this measure can be considered simultaneous to all measuring observers who do not share a reference frame. The inversion of the temporal order due to simultaneity
is impossible to determine, the attempt to measure this effect will unavoidably introduce an uncertainty in the result.
There is no need to have any preferred reference frame for the wavefunction collapse of entangled states.
If an experimenter tries to determine the exact reference frame where the wavefunction collapsed, the measurement process will unavoidably introduce an uncertainty that would make impossible the identification of the ``exact'' reference frame. Obviously, if one can determine the order of the measurement in a shared reference frame it can result like that in certain reference frames and, instead, indeterminate in the other reference frames \cite{resconi,olson}.

\section{Discussion and Conclusions}
It is one of the great merits of Albert Einstein to have investigated the possibility of a multiple-connected space--time and in theoretical physics there is a long tradition of studying quantum behavior in spaces of this type \cite{ho}. 
These lines of research have progressively merged into the quantum study of wormholes, assuming a decisive relevance not only for the study of the structure of the GR and its cosmological implications, but has given the question a decisive configuration as regards the relations of "coexistence" peaceful "between QM and GR. In this work we proposed a formal technique for the study of the quantum effects of a wormhole within the conjecture ER = EPR. We then considered different scenarios from the original Susskind and Maldacena one, in particular those related to the dS space, which seems to be a much more promising ground for the study of the emergence of classical information starting from a quantum background where time is not defined~\cite{vistarini,rovelli2,chiatti,qi}.

These reflections suggest that an effective generalization of the physical meaning of ER = EPR requires a different and more complex philosophy on the emergence of physical space--time as a holographic ``settlement'' of temperature/energy scales, and the use of well-known techniques in QFT~\cite{barvisnky,lisi}.

In other words, these scenarios suggest that the idea of transition of the metric suggested by Sacharov may be the most ``natural'' way to characterize non-locality in a metric formalism.
The~assumption of ER = EPR would be only one of the aspects of a more general phenomenon of Raum--Zeit--Materie production starting from a non-local Euclidean background through quantum computation procedures. The observable part of space time would therefore, in a rather literal sense, result in a thin layer of ice emerging from an ocean of non-locality and the extension of ER = EPR conjecture to Euclidean non-locality may extend its domain from the original AdS/CFT scenario.

Finally, we suggest readers consider the conjecture ER = EPR within the scenario of de Sitter's projective cosmology, described by Hartle-Hawking boundary conditions as Nucleation by Sitter Vacuum \cite{cordafeleppa}. In this cosmological approach one can define the localization conditions in time of the particles starting from an Euclidean pre-space that models a non-local phase. Using the Bekenstein relation, it is possible to identify the area of the micro-horizon $A = (c\theta_0)^2 \simeq 10^{-26}$ cm$^2$, where theta is the chronon, chosen as time scale of the baryonic location. In this case the construction of wormholes applies to a scale much larger than the Planck length. In this case the wormholes are defined by a transition of the metric similar to that hypothesized in the classical work by Sacharov in 1984 \cite{sacharov}.
 
Anyway, the wormhole structure of space--time could in principle be characterized by the extended Heisenberg principle through a deep study of the wavefunction collapse of entangled particles and reveal possible scenarios of QG and cosmology or emergent gravity theories where the exchange of a virtual graviton could also be interpreted in terms of entanglement. At Planck scales wormhole connections would avoid the gravitational collapse and singularities. Moreover, the exchange of a virtual graviton would become equivalent to a wormhole connection and/or entanglement between two or more events. From this we can argue that the ER = EPR conjecture alone, as it is, cannot fully explain without experimental results whether Planck-scale phenomenology can be revealed through entanglement or that gravity and space--time are emergent physical quantities.






\acknowledgments
We dedicate this work to Dino and Gep. F.T. acknowledges ZKM and Peter Weibel for the financial support.

\end{document}